# First observation of temperature dependent lightinduced response of Ge$_{25}$As$_{10}$Se$_{65}$ thin films

Pritam Khan,[1] Rituraj Sharma,[1] Uday Deshpande,[2] and K. V. Adarsh,[1]*

[1]Department of Physics, Indian institute of Science Education & Research, Bhopal 462023, India
[2] UGC-DAE Consortium for Scientific Research, University Campus, Khandwa Road, Indore 452017, India
*corresponding author: adarsh@iiserb.ac.in

Ge rich ternary chalcogenide glasses (ChG) exhibit photobleaching (PB) when illuminated with bandgap light and such an effect is originating from the combined effect of intrinsic structural changes and photo-oxidation. In a sharp contradict to these previous observations, in this letter, we demonstrate for the first time that Ge rich Ge$_{25}$As$_{10}$Se$_{65}$ ChG thin films exhibit photodarkening (PD) at 20 K and PB at 300 and 420 K for continuous illumination of ~ 3 hours. Strikingly, the temporal evolution of PD/PB show distinct characteristics at the temperatures of illumination and provide valuable information on the light induced structural changes. Further, structure specific far infrared (FIR) absorption measurements give direct evidence of different structural units involved in PD/PB at the contrasting temperatures. By comparing the lightinduced effects in vacuum and air, we conclude that intrinsic structural changes dominate over photo-oxidation in the observed PB in Ge$_{25}$As$_{10}$Se$_{65}$ ChG thin films.

The ability to shift the optical absorption of ChGs by bandgap/sub-bandgap light [1, 2] makes them a versatile platform for many potential applications in optics and optoelectronics [3]. To name a few are, fabrication of waveguides that can be used to couple light to various elements in a photonic chip [4], surface relief grating structures as optical filters [5] etc. Among these, ternary Ge-As-Se ChGs are of specific interest because they form a broad composition range of glass forming region and can be easily engineered to observe: photodarkening (PD, optical absorption shift towards longer wavelength) in Ge deficient glasses [6]; photobleaching (PB, optical absorption shift towards shorter wavelength) in Ge rich glasses [7]. In short, composition dependent light induced response of these films indicate that there exist a strong competition between two parallel mechanisms viz. rapid PD and a slower PB. Further, a photostable regime of these glasses is found by choosing the appropriate amount of Ge and As [8, 9].

In recent experiments, we have demonstrated that there exist an unusual coexistence of faster PD (in the initial stages of illumination) followed by a slower PB in Ge$_{25}$As$_{10}$Se$_{65}$ thin films, when illuminated with 532 nm light at room temperature [10]. However, it is not clear how these two opposite photoeffects respond to light at low and high temperatures. Such studies are very important because they will provide new insights on controlling the net lightinduced response as well as the kinetics in ChGs by using a simple tool – temperature of illumination.

On the basis of Raman spectroscopic studies, it is proposed that increase in AsSe$_{3/2}$ pyramidal unit upon illumination gives rise to PD [10]. On the otherhand, a number of investigations have shown that increase in Ge-Se bond density and surface photo-oxidation is responsible for PB [10, 11]. However, the exact assessment on the relative contribution of two processes giving rise to PB is still missing.

In this letter, we demonstrate for the first time that in Ge rich Ge$_{25}$As$_{10}$Se$_{65}$ ChG thin films, in contrast to PB shown at room temperature, shows PD at 20 K Temperature dependent study indicates that kinetics of both PD and PB becomes significantly faster at higher temperatures. A comparative investigations of lightinduced effects in air and in air reveals that intrinsic structural changes dominates over photo-oxidation for the observed PB in a-Ge$_{25}$As$_{10}$Se$_{65}$ thin films. Moreover, structure specific far infrared (FIR) absorption measurements give direct evidence of different structural units responsible for PD/PB at the contrasting temperatures.

The bulk sample of Ge$_{25}$As$_{10}$Se$_{65}$ glass was prepared by the melt quenching method starting with 5N pure Ge, As and Se powders. The cast sample was used as the source material for depositing thin films of thickness ~ 1µm on a microscopic glass substrate by thermal evaporation in a vacuum of 5 × 10$^{-6}$ Torr. The deposition rate was kept below 5 A$^o$/s, which ensured that the composition of the thin films was close to that of the bulk and later confirmed with the EDAX measurements (weight percentage of Ge, As and Se are 24.6, 10.7 and 64.7 respectively). To study the lightinduced effects, at first we have calculated the bandgap of the sample using Tauc plot and is found to be 2.04 eV (608 nm). PD/PB in this film is studied by a pump probe optical absorption method by using the experimental setup described elsewhere [12]. In our experiments, we have chosen the wavelength of the pump beam as 532 nm (above bandgap of the sample) from a diode pumped solid state laser with an intensity of 0.5 W/cm$^2$. The intensity of pump beam was kept constant for illumination at all temperatures. Importantly, the glass transition temperature (T$_g$) of the film is 578 K [13] which is far away from all measuring temperatures. The probe beam was a low intensity white light (450-1000 nm) and its change in transmission before, during and after pump beam illumination was recorded using a high resolution optical absorption spectrometer (Ocean Optics HR 2000) in an interval of one minute. For all experiments, the sample was kept inside an optical cryostat (scan range from 10K to 450K) and the experiments were performed after stabilizing the desired temperature. Far infrared (FIR) pump probe experiments were performed only at room temperature with similar set up using Perkin Elmer Spectrum BX spectrophotometer. The wavelength of the probe beam was in the range of 100 to 400 cm$^{-1}$. For FIR measurements, thin films were coated on the polyethylene substrate which shows good transparency in the FIR region.

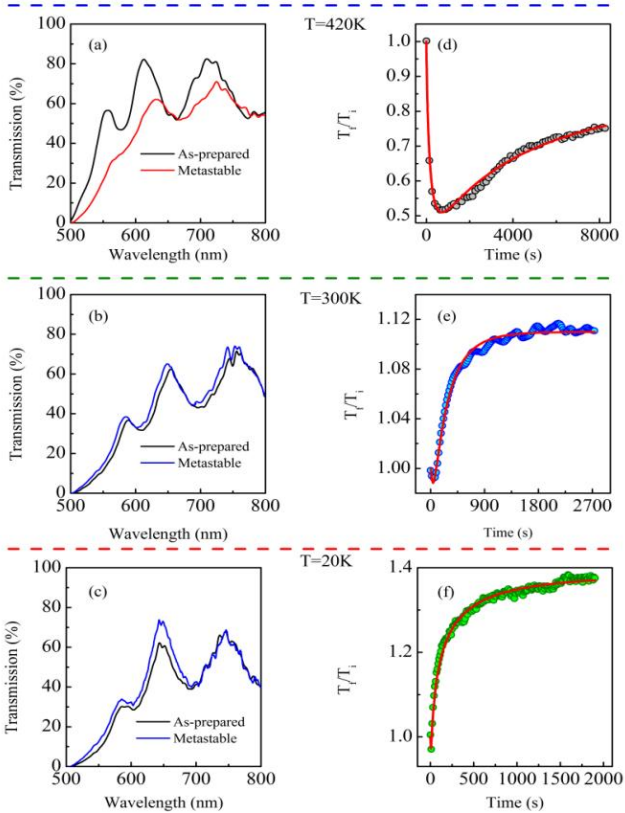

Fig. 1(a)-(c) represent the transmission spectra of a-$Ge_{25}As_{10}Se_{65}$ thin film in as-prepared and post-illuminated state at 20, 300 and 420 K respectively depicting the net lightinduced response. 1(d)-(f) represent temporal evolution of $T_f/T_i$ for the wavelength at which initial (dark) transmission is 10% at the respective temperatures. The solid red lines in fig. 1(d)-(f) represents the theoretical fit of experimental data.

Fig. 1 shows the net light induced response of a-$Ge_{25}As_{10}Se_{65}$ thin films at 3 different (20, 300 and 400 K) temperatures. Pump beam illumination produces a net PD at 20 K after continuous illumination of ~ 3 hours, however it switched to a net PB at 300 and 400 K (fig. 1a-c). At this stage we envision that the crossover from PD to PB is a function of temperature that pass through an intermediate temperature, where the sample is photostable i.e. it shows resistant to light induced changes. This experimental observation is of great importance since it gives direct evidence that temperature can be used as an effective tool to control light induced response and thereby achieving photostability.

To get deeper understanding on the kinetics of lightinduced effects, we have analyzed the temporal evolution of $T_f/T_i$ at probe wavelengths for which transmission was 10% of the value in the dark or as-prepared condition, at those temperatures and is shown in fig. 1d-f. At 20K, $T_f/T_i$ decreases instantaneously, an observation consistent with PD and saturates within a few minutes. After that, $T_f/T_i$ starts increasing that is in consistent with PB, and ended up at a value well below the as-prepared state to produce net PD (fig. 1d). At 300K, PD begins almost immediately upon illumination and saturates within a few tens of seconds with a much lower magnitude than 20 K. Subsequently, PB starts to grow and saturates at a value which is well above the initial transmission, producing a net PB (fig. 1e). Light induced effects at 420K (fig. 1(f) show dramatically distinct characteristics when compared to 20 and 300 K. In this temperature, we could observe only negligible PD and PB start to dominate from the very beginning. Further, the magnitude of PB is much higher than the room temperature values.

In our next step, to model the reaction kinetics of opposite lightinduced effects (PD and PB) at three contrasting temperatures, we have used a combination of stretched exponential functions that describe PD and PB separately [10]:

$$\Delta T = A[\exp\{-(t/\tau)^{\beta_d}\}] + \Delta T_{Sd} + \Delta T_{Sb}[1 - \exp\{-(t/\tau)^{\beta_b}\}] \quad (1)$$

Here the subscripts 'd' and 'b' correspond to PD and PB, respectively. $\Delta T_S$, $\tau$, $\beta$, t and A are metastable part, effective time constant, dispersion parameter and illumination time and a temperature dependent quantity which is equal to the maximum transient changes respectively. We have theoretically fitted our experimental data using Eq. (1), which fit very well — see Fig. 1(d)-(f). Fitting parameters calculated from theoretical fit are listed in Table 1.

Table 1. Time constants and dispersion parameters obtained from Eq. 1 that corresponds to PD and PB for a-$Ge_{25}As_{10}Se_{65}$ thin films at different temperatures.

| T(K) (illuminated in vacuum with negligible oxygen concentration) | $\tau_d$ (s) | $\beta_d$ | $\tau_b$ (s) | $\beta_b$ |
|---|---|---|---|---|
| 20 | 170 | 0.80 | 3600 | 0.82 |
| 300 | 39 | 0.92 | 210 | 0.81 |
| 420 | 3 | 0.71 | 85 | 0.78 |

From the table we could observe two most important features:

(1) Kinetics of both PD and PB become dramatically faster as temperature increases. We could observe a

significant decrease in both $\tau_d$ (from 170 seconds to 3 seconds) and $\tau_b$ (3600 seconds to 85 seconds) when illumination temperature is raised from 20 to 420 K. Such observations clearly indicate thermal vibration accelerates the kinetics of PD and PB at higher temperatures. (2) At any measuring temperatures, kinetics of PD is one order faster than that of PB (at 20 K, $\tau_d$ is 170 seconds, whereas $\tau_b$ is 3600 seconds, from Table 1) — a clear indication that PD is an instantaneous process having very short reaction time whereas PB needs some latent time to start with longer reaction times.

As our next step, we will show that the light-induced change consist of a transient (which persists only during illumination) and a metastable component. Fig. 2(a) shows the time evolution of transient part between "on" and "off" states of laser illumination at three contrasting temperatures.

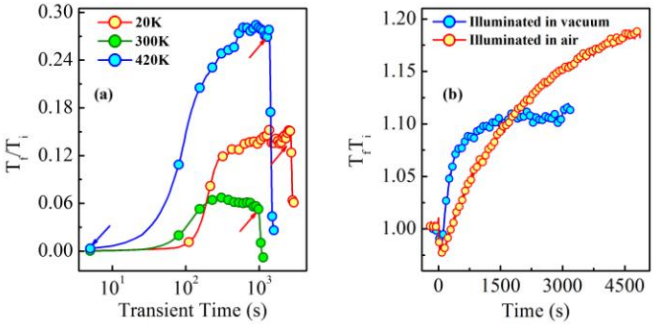

Fig. 2. (a) Time variation of transient component of lightinduced effects at 20, 300 and 420K between on and off states of laser illumination. The downward (blue color) and upward (red color) arrows represent the time when laser is turned off and on respectively. (b) Comparative lightinduced effects in a-$Ge_{25}As_{10}Se_{65}$ thin films when illuminated in vacuum and air.

Comparing the magnitude of transient PD (TPD) at three contrasting temperatures, we found that TPD is largest at 420 K. Notably, transient effects become negligibly small at room temperature. Surprisingly, sample exhibits appreciable TPD at low temperature which is intermediate to the value observed at room temperature and 420 K.

To explain the observed effects, PD occurs because of the reduction in As homopolar bonds presumably present in As-rich clusters in the following way [1, 9]:

$$As_{(x-y)}Se_z + yAs + h\nu \leftrightarrow As_xSe_z \quad (2)$$

However, for PB two parallel mechanisms are proposed. First, intrinsic structural changes described by the following equation [1]:

$$Ge-Ge + Se-Se + h\nu \leftrightarrow 2Ge-Se \quad (3)$$

Second, photo-oxidation of Ge atoms by the creation of Ge-O bonds at the expense of Ge-Ge bonds that are broken by illumination [14].

Therefore, to quantify directly the contribution of intrinsic structural change and photo-oxidation in determining PB, we have performed in-situ pump-probe measurements both in air in and vacuum (where the oxygen concentration is negligible). For measurements at the vacuum the sample was kept inside an optical cryostat evacuated to $2 \times 10^{-6}$ Torr. In this context, figure 2(c) show the time variation of $T_f/T_i$ of a-$Ge_{25}As_{10}Se_{65}$ thin films when illuminated in vacuum and air respectively. It can be seen from the figure that upon illumination, in both the conditions, sample undergoes instantaneous PD followed by a slower PB. However, we could observe that magnitude of PB is higher in air compared to vacuum. Moreover, saturation time for PB is found to be longer in air in comparison to vacuum. In order to quantify the effects, we have calculated various parameters as described previously [10], which are summarized in table 2.

Table 2. Calculated parameters of a-$Ge_{25}As_{10}Se_{65}$ thin films when illuminated in air and in vacuum (negligible oxygen).

| | $\Delta T_{PD}$ | $\Delta T_{PB}$ | $\tau_d$ (sec) | $\tau_b$ (sec) |
|---|---|---|---|---|
| Illuminated in vacuum | 0.02 | 0.13 | 39 | 210 |
| Illuminated in air | 0.02 | 0.21 | 38 | 1896 |

By looking at the Table 2 carefully, we get new insights on lightinduced effects as:

(1) Both magnitude and kinetics of PD remains invariant in both the conditions indicate that intrinsic structural rearrangement is sole mechanism responsible for PD.

(2) Magnitude of PB in air ($\Delta_{PB}$=0.21) is much larger compared to vacuum ($\Delta_{PB}$=0.13). Precisely, ($\Delta_{PB}$) vacuum equals 62% of ($\Delta_{PB}$) air. From this, we conclude that the intrinsic structural changes dominate over photo-oxidation for the observed PB in a-$Ge_{25}As_{10}Se_{65}$ thin films.

(3) Kinetics of PB in vacuum is much faster compared to that of air. Reaction times of PB in air ($\tau_{air}$=1896 seconds) is found to be ten times slower than in vacuum ($\tau_{vacuum}$=210 seconds) with negligible oxygen concentration. This is a consequence of the fact that surface photo-oxidation of Ge atoms is a slower process compared to intrinsic structural rearrangement which makes the kinetics of PB drastically slower in air in comparison to vacuum where the possibility of photo-oxidation is very negligible.

To get deeper understanding on the origin of lightinduced effects we have also performed FIR absorption spectra of a-$Ge_{25}As_{10}Se_{65}$ thin films in as-prepared and illuminated state and are shown in figure 3. It is quite evident from the figure that dominant features of the FIR spectra of a-$Ge_{25}As_{10}Se_{65}$ thin films consist of three independent modes: (1) a weak peak ($M_1$) centered at 237cm$^{-1}$ is attributed to the $v_7$ modes of As-Se structural unit and a contribution from $A_1$ and E modes of Se polymeric chain [15, 16] (2) a strong peak ($M_2$) occurs at 256cm$^{-1}$ is assigned to the asymmetric bond-stretching mode $F_2$ of $GeSe_{4/2}$ tetrahedra [17] and (3) a broad peak ($M_3$) centered at 307cm$^{-1}$ is identified as the Raman mode of $GeSe_2$ [16, 18]. To understand the underlying mechanism of PD/PB, we will now analyze the FIR data in more detail and calculate the absorption peak ratio of different vibrational modes. In doing so, we have normalized the data with respect to the most intense peak $M_2$. The results indicate that, with illumination $M_2/M_1$ decreases from 1.88 to 1.35. This observation shows that density of $GeSe_{4/2}$ tetrahedra decreases by light illumination whereas density of As-Se structural unit, Se polymeric chain increases which together gives rise to PD. On the otherhand, considering the peak $M_3$, we found that $M_2/M_3$ also decreases with light illumination. $M_2/M_3$ was found to be 3.84 in the as-prepared state, which becomes 2.22 in the illuminated state. This result points out that upon illumination density of $GeSe_2$ Raman mode increases which strongly favors PB. Therefore, with light illumination, there exist two competitive mechanisms, viz. PD arising because of increase in Se chain and As-Se structural units and PB through the increase in $GeSe_2$. By comparison, we found that effective change in $M_2/M_3$ ($\Delta M_2/M_3 = M_2/M_3$ before illumination – $M_2/M_3$ after illumination=1.62) is more than the effective change in $\Delta M_2/M_1$=0.53 which indicates that light illumination causes more increment in the density of $GeSe_2$ compared to As-Se structural unit and Se polymeric chains. Therefore, PB dominates over PD in the net lightinduced response for a-$Ge_{25}As_{10}Se_{65}$ thin film.

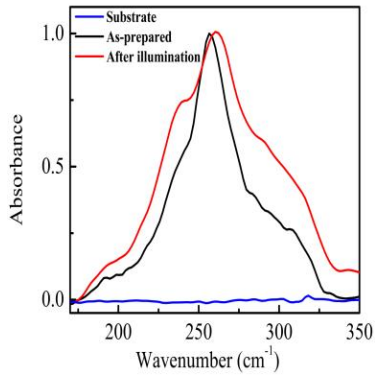

Fig. 3. FIR spectra of as-prepared and illuminated a-$Ge_{25}As_{10}Se_{65}$ thin films.

In conclusion, we have demonstrated for the first time temperature dependent lightinduced response in a-$Ge_{25}As_{10}Se_{65}$ thin films: PD at low temperatures and PB at room and higher temperatures. In addition to this, we could observe a dramatic influence of temperature on the kinetics of both PD/PB—both the effects are significantly faster at higher temperatures. Enhancement of $GeSe_2$ over As-Se structural units and Se polymeric chains gives rise to net PB. Finally, a comparative study of lightinduced effects in vacuum with negligible oxygen concentration and air clearly demonstrates that the intrinsic structural changes are more dominant than photo-oxidation in the observed PB in these films.

The authors thank Department of Science and Technology (Project no: SR/S2/LOP-003/2010) and council of Scientific and Industrial Research, India, (grant No. 03(1250)/12/EMR-II) for financial support. We also thank UGC-DAE consortium for scientific research for FIR measurements.